\documentclass[11pt]{article}

\usepackage[T1]{fontenc}
\usepackage[utf8]{inputenc}
\usepackage{lmodern}
\usepackage[margin=1in]{geometry}
\usepackage{microtype}
\usepackage{amsmath,amssymb}
\usepackage{enumitem}
\usepackage{booktabs}
\usepackage{array}
\usepackage[authoryear,round]{natbib}
\usepackage{hyperref}
\usepackage{url}
\usepackage{verbatim}

\hypersetup{
  colorlinks=true,
  linkcolor=blue,
  urlcolor=blue,
  citecolor=blue,
  pdftitle={The Graded Kilobyte: Assembling a Commodity Market for AI Inference},
  pdfauthor={Nicholas R. Sarantinos}
}

\setlist[itemize]{itemsep=2pt,topsep=2pt}
\setlist[enumerate]{itemsep=2pt,topsep=2pt}

\title{The Graded Kilobyte:\\Assembling a Commodity Market for AI Inference}
\author{%
  \textbf{Nicholas R. Sarantinos}\\
  Veltisto\\
  \href{mailto:nicholas@veltisto.ai}{\textcolor{black}{\texttt{nicholas@veltisto.ai}}}
}

\date{}

\begin{document}

\maketitle

\begin{abstract}
Today AI inference is sold in heterogeneous, hard-to-compare units that are not well aligned with buyers' incentives. It is also largely a credence good: buyers cannot reliably judge quality even after the answer arrives. We argue that solving these two problems is enough for a commodity market in AI inference to assemble itself. For measurement, we propose a \textbf{meter}: a price per kilobyte of zstd-compressed output, with input and reasoning unmetered. Anyone can reproduce the measurement, and padding discounts itself. The meter counts volume but never judges, because a meter that judges is a meter that can be argued with. For quality, we propose an \textbf{expert}: an institution that grades models and sellers through procedurally generated benchmarks with no fixed test set that can leak, live probes with known answers mixed into production traffic, and routing informed by both. On this foundation, a market assembles itself. The product is the \textbf{Graded Kilobyte}: output certified statistically rather than by inspection, like a tanker of tested milk, and sold in grades (100, 97, 85) that are re-marked as the frontier moves. On top of that: a live, quality-adjusted index. Forwards and full-requirements service. Then auctions, speculation, and finally an exchange whose forward curve prices the pace of AI progress. Safety is the same credence problem, worse: the public debates it without hard measurements to check. The expert publishes third-party safety measurements. Finally, this new market can inherit a century of commodity regulation rather than wait for new law.
\end{abstract}

\section{Introduction}

A century ago, every factory that wanted electricity built its own power plant. Then the market arrived: spot prices, off-peak windows, forward contracts, full-requirements service, and a new profession whose edge was knowing when and where to buy. The factories that stayed on the retail tariff ended up subsidizing everyone else's efficiency.

We argue that a market for AI inference has two prerequisites, propose them, and trace the market that follows. The first is a meter both sides can read, and the second an institution of quality both sides can trust. The meter made volume a fact. The institution made quality something you could underwrite. Everything else-the tariffs, the futures, the exchange, even the regulation-followed from those two inventions.

\textbf{The first prerequisite is a meter aligned with the buyer's incentives.} The forgotten chapter of electricity's history is the billing one. Early lighting was priced per lamp per month-the only quantity buyer and seller could count without instruments. Lamps burned whether anyone needed light or not, and cities wrote \textbf{``moonlight clauses''} into their contracts: lights off under a full moon. When you cannot meter what you sell, you ration it with fine print. Around 1889, the recording wattmeter shifted billing to the kilowatt-hour delivered, and per-lamp pricing died within a decade. Volume became what economists call a \textbf{search good} \citep{nelson1970}: observable before purchase, verifiable after.

Notice what the watt-hour meter did not measure. Not brightness. Not usefulness. Not whether anyone was in the room. It measured one thing-volume, an exchangeable quantity both sides could agree on. Reliability, voltage stability, the competence of the utility: all handled separately, by reputation, contract, and competition. The meter was not the market. It was the one piece of the market that required no trust. That was enough.

Inference today is where electricity was in 1889. The model is becoming the commodity, the providers are the generators, the bill is the tariff-and most buyers still pay the retail rate, routing every prompt to the most expensive model. The reason is structural. Inference is largely a \textbf{credence good} \citep{darby1973}: even after the answer arrives, the buyer often cannot distinguish subtly wrong code from correct, or plausible-but-false reasoning from sound. Worse, a prompt's difficulty is unobservable until the answer exists-inference difficulty cannot be observed without being served. The formal results on credence goods are unforgiving: buyers rationally default to the expensive option \citep{dulleck2006}. Paying for a frontier model on every prompt is like buying zero-deductible insurance on every household item, down to the coffee mugs.

An enterprise that cannot certify the quality of what it buys does not buy much of it. Every team that routes a trivial summarization to a frontier model learns the wrong lesson-that inference is expensive-and shelves the hundred deployments that would have cost a tenth as much. Judging the frontier by what it charges to summarize a paragraph is judging the fire department by the price of rescuing a cat from a tree. The retail tariff is a ceiling on adoption, and it binds the frontier sellers hardest.

\textbf{The second prerequisite is the standard cure for a credence good: certification.} A middleman-as-expert screens quality better than dispersed buyers can \citep{akerlof1970,biglaiser1993,lizzeri1999}. The alternative is for every team to become its own model picker-a job none of them signed up for.

The meter counts volume. The expert grades quality. Each does the job the other cannot. The rest of this proposal builds each in turn, then watches the market assemble around them.

\section{The Meter}

\subsection{The Search for the Inference Watt-Hour}

Today inference is metered in input tokens, output tokens, cache tiers, batch rates, and off-peak rates-workable for engineers, unusable for a market. We began the search for the inference watt-hour with a disciplined definition: a measure of volume delivered, and nothing more.

The candidates, and why they failed:

\begin{itemize}
  \item \textbf{Compute.} Estimated FLOPs per prompt cannot be verified by the buyer. A volume measure works only if both sides can check the ruler.
  \item \textbf{Unified token pricing.} Tokens are not provider-independent: counting them requires an API call on some providers, and tokenizers vary across models.
  \item \textbf{A learned quality estimator.} Rejected for a counterintuitive reason: it is not a volume measure at all. The moment a meter judges, the judgment becomes contestable, and the contest is the seller's opening. A meter that judges is a meter that can be argued with.
  \item \textbf{Per-request pricing.} This taxes the one direction inference must evolve: autonomous work. The future is one prompt that sets a model loose for hours to build a website or migrate a codebase. Under request billing, the seller does best by stopping early and splitting the task into billable checkpoints. The buyer's best outcome-a finished artifact delivered in one run-becomes the seller's worst.
\end{itemize}

Output bytes survived. There is precedent for this kind of search: units of trade are chosen to keep measurement cheap, not to match physical value \citep{barzel1982}. Coal sold by the cart until the calorimeter made heat content cheap to verify. Only then did the Btu market appear. Token billing is Barzel's world-a unit chosen for the seller's convenience. We set out to choose one for the buyer's.

Then the key insight: charging for input is charging for fuel burned, heat losses included, instead of energy delivered. The buyer does not care how the model works under the hood. A recurrent network and a transformer should bill identically for identical output. So we dropped input from pricing-and reasoning with it. Generation losses belong to the generator.

This has to survive contact with the seller's cost sheet. Generation is the cost: each output token attends over everything before it, so cost grows with context. Input billing was a defensible proxy for that deepening attention bill-but the proxy is dissolving under prompt caching and sparse and linear attention. Ours is a bet on where the cost structure settles, not where it stands today.

Flat pricing has been tried and abandoned. OpenAI's GPT-3 models each billed a single token rate from the 2020 API launch. The split came with GPT-4 in 2023. Mistral did the same. But flat pricing failed because a blended rate hides more than a split does. Output-only hides nothing: it charges only for what has value to the buyer. It is the most legible tariff, not the least.

The edge cases are real. A hundred-thousand-token prompt that yields one word is expensive to serve and undercharged. The gap closes from three directions.

First, engineering. Prompt caching serves repeated context at a fraction of face cost-and under output-only billing the seller keeps the entire saving. If the seller is an expert too, context compaction narrows the gap further: methods like LLMLingua-2 \citep{pan2024} strip a bloated prompt before it ever reaches the model. Published results leave wide headroom-headroom worth an expert's research budget.

Second, the buyer's own incentive. Models perform worse when long context is padded with irrelevant material, so a buyer who trims a prompt buys a better answer. Where quality already punishes bloat, the meter need not.

Third, price. Today input is billed above the cost it incurs. Output-only billing deletes the explicit input charge, and the market prices what remains: providers fold the true cost of input into the rate they quote for output. If the rate fails to cover costs, the rate rises, and the market adapts. The seller prices the portfolio, not the prompt. An expert that buys input, output, and reasoning tokens and sells kilobytes of delivered output is an insurer-and no insurer prices the individual claim. It prices the pool. A rate that clears in aggregate is a rate that holds. (Adversarial behavior is the expert's to take up.)

Cost accounting tells the seller what it needs. The meter tells the buyer what it gets. These have always been different numbers-a power plant's coal ledger and a customer's meter never agreed either.

Output-only pricing also hands the frontier labs a gift. Under token billing, an efficiency gain that halves reasoning cost shrinks the billing base-the replacement effect \citep{arrow1962}, the reason incumbents have little incentive to fund inventions that cannibalize their own revenue. Under our meter, reasoning is unmetered: the lab delivers the same kilobytes at the same price and keeps the entire saving as margin. With quality held at a certified standard, a rival cannot undercut without first matching the efficiency. The meter does not just price fairly-it funds the next architecture.

\subsection{The Compressed Kilobyte}

One attack remained: padding. A seller can grow measured volume with verbosity the buyer never wanted, and a meter that rewards dilution measures the seller's incentive, not the buyer's. That left exactly one move: make padding discount itself.

\textbf{Charge per kilobyte of zstd-compressed output. Input and reasoning unmetered.}

One command reproduces any measurement: divide the printed byte count by 1,000, then multiply by the price posted at the start of the request.

\begin{quote}
\texttt{zstd -c your\_response.txt | wc -c}
\end{quote}

Default options, pinned to version 1.5.7, chosen over gzip for its ratio-speed balance. (Disconnections are a non-issue: the zstd frame header means even a truncated stream bills nonzero.) The public standard is a necessity, not a courtesy. A proprietary yardstick unravels no one. A verifiable one forces the whole market to follow, because once one seller discloses credibly, non-disclosure becomes informative \citep{milgrom1981}. A verifiable meter doesn't just let buyers check invoices-it makes every competitor's opacity read as concealment.

Can the seller game it? Raising entropy-quietly running temperature 1.2 against a requested 1.0-fattens output only enough to be noticed. Inference is not entirely a credence good, and artifacts show. At the other extreme, repetitive garbage compresses to near zero and doesn't pay. A model built from scratch to inject plausible high-entropy filler faces the same bind: if the filler is relevant it compresses, and if it is not, the buyer notices. Whatever slips past the buyer is precisely what the expert exists to catch, by watching the distribution of compressed output.

The division of labor has precedent. A taxi meter measures distance, not driving skill. The passenger screens the driver separately. A watt-hour meter measures volume. The grid guarantees voltage separately. Ours is the same architecture: the meter prices volume, the expert gates quality.

The sharpest objection deserves its answer: electricity is fungible and inference is not. A kilowatt-hour is a kilowatt-hour. A compressed kilobyte may be insight or confident nonsense. True-and that is the point, not a flaw. The watt-hour worked with thin quality institutions because electricity itself needed none: every electron performs. Because the kilobyte is not fungible, inference needs a thick one. The non-fungibility of text is the reason the expert must exist, and the two-layer structure keeps each half honest.

\section{The Expert}

The meter counts volume. It cannot judge quality, and we have argued it should never try. But a market that prices volume without measuring quality rewards dilution. Something must stand between the buyer and the question: is this answer any good?

That something is the expert. Electricity itself never needed one-every electron performs. The devices built around it were another matter. The safety of wiring inside walls was a credence good: no homeowner could tell a safe installation from a fire hazard by looking, and insurers refused to underwrite buildings full of untested ones. The underwriters' answer, in 1894, was a laboratory of their own-Underwriters Laboratories-whose mark made a credence good purchasable. Our expert is that laboratory, with one difference of scope: a toaster's design, once certified, stayed certified. A model's grade is true only until the model retrains. The expert must grade models as products, sellers as operators, and live traffic as it flows-continuously, in a market that renews itself every few months. No single method suffices. The expert is an architecture of methods, each covering the failure modes of the others.

\subsection{Static and Live Grading}

\textbf{Static grading.} Public benchmarks saturate on a treadmill for one reason: models train on them, often unintentionally, as they circulate. A public benchmark is a test whose answer key leaks the moment it is published. The fix is \textbf{procedural generation}: task environments-invented mini-languages, synthetic APIs, novel rule systems-where each instance is produced from a private seed at evaluation time. Contamination becomes structurally impossible. There is no dataset to leak, because the dataset does not exist until the moment of the test.

The generator yields a second instrument for free. For each task, produce a canonical variant a model may have memorized and a novel variant that is equally hard for a human. The score is the delta. A model at 85\% canonical and 82\% novel has real skill. One at 90\% canonical and 40\% novel is contaminated. The novel variant measures what a model can do with what it hasn't seen. How procedurally generated worlds map onto real expertise is the hard, proprietary core of the institution, and a strategic expert keeps it to itself.

One property shapes the economics of everything that follows: a diamond's grade is true forever. A model's grade is true until the model trains on the grader. When a measure becomes a target, it ceases to measure \citep{goodhart1975}-and here the gaming is adversarial \citep{manheim2018}, because the measured are machines that re-arm with every training run. Certification is a \textbf{Red Queen expense} \citep{dawkins1979}: you keep paying to stay in place. \citet{grossman1980} showed that information must earn at least what it costs to gather, or no one gathers it. Our variant is harsher: the signal is destroyed not by nature but by suppliers training against the ruler.

The expert's costs therefore split into two ledgers. A \textbf{pass-through ledger}: upstream prices times realized dispatch. A \textbf{measurement ledger}: evaluation compute, held-out regeneration, adversarial re-grading. The ledgers diverge. Producing answers gets cheap. Distinguishing them does not. The closer two models are, the more testing it takes to tell them apart. As the frontier converges, the gaps that matter shrink, and the expert's bill grows. The bottleneck migrates from compute to research, and insight into where machines fail rides no cost curve. National metrology institutes have never shrunk for this reason. The expert's rent persists, indexed to the capability it measures.

\textbf{Live grading.} Static grades go stale the moment a model ships. Sellers quantize under cost pressure, hotfix mid-week, and inherit upstream changes-none of it visible to any pre-release benchmark. So the expert audits the stream itself. Passively: compression ratios from the meter, linguistic features of inputs, similarity matching against known responses-cheap, continuous, weak in isolation. Actively: \textbf{probes}-a strategically chosen fraction of requests drawn from the procedural generators, whose ground truths are known in advance, so the answers grade themselves. Baseline probing can run at 0.05\% of traffic, escalating only on suspicion. With enough volume, oversight is a rounding error.

Probes inherit the generators' structural immunity: the seller cannot learn to recognize them. A seller could attack the expert by leaking probes to weaken its authority-but only anonymously, and a strategic expert watermarks probes with seller-identifying information. The leak evaporates the seller's reputation, not the expert's. And drift requires no bad faith: probes catch it whether it originates in a decision or a dependency, and they exonerate the honest provider whose numbers were fine all along. The honest seller has as much stake in the audit as the buyer. Alongside quality, the expert grades latency, throughput, and disconnect rates, builds each seller a reputational profile-and acts on it: reward the top, exclude the bottom.

\textbf{The judgment layer.} Procedural generation can supply correct answers only for tasks with definite solutions, such as certain problems in mathematics or programming. A legal memo and a business recommendation have no single right answer. The usual fallback is to use one or more off-the-shelf frontier models as judges. But adding judges does not eliminate bias: LLM judges can favor their own outputs and those of related models, while familiarity and rubric design can distort their scores \citep{panickssery2024,wataoka2025}.

A number of methods can reduce these biases. The strongest is a judge the expert builds itself: a purpose-built model trained on a large proprietary dataset of response comparisons made by subject-matter experts against explicit criteria. The dataset must span the tasks, failure modes, and standards the judge is meant to handle. Blinded tests on unseen examples and adversarial cases can measure its reliability while periodic recalibration keeps it current. The judge need not be infallible to be authoritative: its uncertainty is measured. Its judgments support statistical grades of models and sellers, not guarantees about individual answers. It is one of the expert's most valuable assets, the instrument that makes open-ended quality gradable at market scale.

\textbf{Safety.} A quality failure eventually surfaces. A safety failure may never. So safety gets a floor, not a grade: a binary attestation. Everything behind the stamp is published-behaviors probed, base rates assumed, detection odds computed-and no lab sees more of the record than the public. What hides in production fires under test: ``may never'' becomes ``likely,'' at computable odds. What was tested, what was found, when a stamp came off. Blind spots included. Underwriters Laboratories never opined about toasters. It stamped a fact about a test. AI safety requires assertions on top of hard facts, but the expert should weight the facts more than the assertions.

\subsection{Routing, and the Question of Trust}

Assigning each prompt to the right seller-model pair is a quality judgment, and it belongs to the institution that grades quality. Without it, the market decouples back into its parts, and the buyer's only rational strategy is the retail tariff: overbuy. A rich routing literature exists \citep{ong2024}, and genuine novelty is still possible-but the moat is not only the method. Anyone can read router papers. Only an expert with proprietary traffic, private benchmarks, and live reputational data can learn which routing signals predict quality in production, at scale, across models that change every few months. Routing also closes the feedback loop: a grade that never touches a routing decision is just a report. A routing decision informed by live probes is a market signal.

This raises the obvious question: who grades the grader? The expert is itself a credence good, cured by the usual institutions: reputation, skin in the game, and competition \citep{klein1981}. Buyers cannot grade individual answers, but they can grade outcomes-whether costs fall, whether overbuying stops-and they can switch. Competition is the real check. The moment a second expert appears, the incumbent's grades become contestable. Rating agencies were graded by the market too, and the market graded too late. The difference here is that experts audit each other for a living. The expert remains the most reputation-fragile institution in the market: a bad utility's lights still turn on. A bad expert's buyers may never find out the quality was below standard.

There is also an open-source variant. Procedural generation separates method from test: the generator architecture can ship while the seeds stay private, as a hash function is published while its nonces are not. Seller-side reputation profiles contain nothing confidential and can be published the way index funds publish holdings-the observations are the product, not the secret. What must stay closed is narrow: buyers' raw queries and outputs (confidential), and the private seeds. Even routing can be disclosed in aggregate, the way exchanges publish order-flow summaries. Such an expert can still be wrong, but it can hardly be accused of hiding-and against a closed rival, that is a weapon. The sequencing matters: open too early and you hand rivals your methods before your data moat exists. Open late, and transparency adds little because the moat has moved from code to observations. The alternative path-taken up later-substitutes regulation for transparency: submit to supervision on an exchange and let the regulator's audit do the trust-building.

\subsection{The Spread}

How is the expert paid? It buys inference upstream at wholesale and quotes buyers a slightly higher rate, keeping the difference. \textbf{The spread is incentive-compatible by construction}: the expert earns only when routed volume clears at a price buyers accept. Serve badly and the revenue stops. The payment itself punishes misbehavior. No oversight required. The buyer pays for a service received-sourcing, routing, certification-never for a verdict. Competition disciplines the spread as it disciplines any market maker: any rival can quote tighter.

Two prohibitions protect the spread's integrity. \textbf{Sellers never pay for their grade.} And the expert, though free to sell forwards, may never carry open exposure to its own index: every forward sold must be closed by a matching purchase upstream. An uncovered forward is not dealing-it is a bet on a number the expert itself writes.

The balance sheet inherits basis risk because the expert buys in input, output, and reasoning tokens and sells in compressed kilobytes. One upstream model update can reprice every forward-sold kilobyte. The standard remedy is a cross-hedge: an offsetting position in a correlated instrument. Every dealer who bore such a gap uncovered learned what a correlation breakdown costs. So an expert that has sold six months of kilobytes forward buys matching capacity forward from qualifying sellers, month for month. Where the match is imperfect, the residual prices visibly into the spread, as any market maker widens quotes on unhedged inventory.

Is the spread legitimate? It is real money over wholesale-but it buys something real. It prices three things the buyer cannot otherwise purchase: certification of a credence good, matching of prompt to seller, and certainty across time. Electricity's earliest utilities earned just such a spread, tolerable for one reason: it was visible, contestable, and shrank the moment a competitor quoted tighter. A middleman's margin is legitimate exactly as long as the market can see it, price it, and route around it.

\section{The Market}

\subsection{The Pre-Market Era}

Before the market, there are marketplaces. Aggregators put sellers side by side, hand every new model a storefront on day one, and have taught buyers to compare. Behind the storefronts, providers reprice daily or weekly under pressure no buyer can see. The ingredients are on the shelf. What is missing is the meter and the expert. Without them, the buyer who sorts by cost per million tokens must still audit quality by hand-the one job nobody signed up for-and ends up comparing tariffs in units of effort rather than delivery: fuel and heat losses included. Sellers sit on idle capacity they cannot sell and throttle at peak. Buyers overpay for quality no one warrants. Every commodity market, electricity included, looked like this on the way to a meter.

\subsection{The Graded Kilobyte}

The meter counts volume. The expert grades quality. Neither touches the oldest problem in any commodity market: tomorrow's price is unknown. The seller has the generator's problem-sunk capital, output that cannot be stored. Price high and lose customers, price low and leave money. The buyer has the industrial buyer's problem-a product budgeted on inference needs next quarter's cost in advance. Both want certainty. Neither can manufacture it. In every commodity market, certainty is manufactured by a third party paid to bear the risk: the speculator \citep{keynes1930,hicks1939}.

What should a derivative on intelligence be written on? Futures settle on standardized, fungible quantities, and inference is not fungible. Every commodity market has solved this not by making the good uniform but by defining a grade: a corn contract is written not on corn but on corn No.~2 Yellow. The expert cannot certify individual answers-quality is verifiable only in aggregate, through probes that sample the stream-but it can define, publish, and stand behind a basket: the stream it continuously audits. That basket is the first genuinely fungible quantity in this market, fungible the way a tanker of tested milk is: statistically, by certification rather than inspection.

So: the \textbf{Graded Kilobyte}-a kilobyte of compressed output, scored Grade-100, Grade-97, Grade-85. Grade-100 (Grade A) marks the frontier: the strongest publicly available model on the grader's instrument, normalized to 100 at each grading run. Grade-97 doesn't mean ``3\% worse'' on any absolute scale. It means 97 index points against a fixed basket of task families-cardinal by construction, because the index defines the aggregation. Grades are re-marked as the frontier moves. That is a feature: a grade that meant the same thing forever would price a decaying asset at par. Composite systems can grade above the model-level frontier-Grade-105 (Grade A+).

The grade must be distinguished sharply from today's tiers. ``Flagship'' and ``mini'' are declared by the seller and defined against no fixed reference. ``Mini'' means ``worse than our flagship,'' not ``worse than the market.'' Tiers are honest answers to a question no seller has the instruments to grade. A grade is declared by an institution paid to be right, defined against a committed frontier, and enforced by a market that prices every deviation from rival graders' verdicts. Choosing between mini and flagship is an act of faith in a brochure. Choosing between grades is an endorsement of one grader among competitors who audit each other for a living.

The grade certifies one attribute-quality-and everything else is contracted around it: privacy terms, data sovereignty, throughput floors, percentile latency targets. Corn futures certify grade while storage, delivery point, and timing are separate terms. The grade stays clean: one number, one meaning.

The ladder needs a price to trade against: \textbf{the index}-the volume-weighted, time-averaged price of graded kilobytes across sellers above threshold, mechanically derivable from grades and metered receipts. It is a hedonic index \citep{griliches1961}-a price index adjusted for quality-with a difference. Hedonic indices have always priced a quality-adjusted past: the BLS computes them retrospectively, years behind. Case-Shiller futures trade but settle quarterly. The Baltic Dry prints daily and trades through forward freight agreements, but prices a homogeneous good. This one is live, quality-adjusted, and settles contracts. We know of no other index that is all three.

Products follow in order. Bilateral forwards: a fixed rate per graded kilobyte for six months, settled against the index-electricity's forward market ran bilaterally for decades. Full-requirements service: the expert covers a buyer's entire workload at one rate, bearing the sourcing risk and standing behind every kilobyte. This is the enterprise product the market has been missing. Electricity's version was reserved for large industrials. This one will be too, at first-and frontier labs are its natural sellers. Then caps for buyers, and the grade ladder extending downward and upward from the frontier as the market matures.

This inverts the economics of commoditization. When smartphones commoditized, the premium players drowned with the middle. Apple and Samsung survived by making quality legible to buyers who could not see it. But their premium rang up only their own products, and had to: licensing iOS to a rival would spend the very scarcity the premium is made of \citep{teece1986}. The expert's premium is made of the opposite ingredient-authority, which grows the more models it touches. That is the step not even Apple could take: a toll that rings up the whole market, collected on value others create. A lab that runs an expert arm profits from its own commoditization: every new model seller is new traffic to grade. Its revenue, once bounded by the models it sold, becomes a claim on the market's entire traffic-including penny-a-task traffic no lab could ever have sold at a profit.

\textbf{The monopoly question.} The expert is no monopolist in the commodity sense. A commodity monopolist controls supply while the expert controls a standard. The law tolerates standards precisely when they are contestable. The grade is a warrant, not a barrier: any seller that clears the published thresholds enters the index, and any buyer that distrusts the incumbent can route around it. The spread invites its own competition. The moment a second expert quotes tighter, the incumbent's margin becomes its vulnerability. The true safeguard is architectural, not legal: the meter is public, the index is derivable from verifiable receipts, and the buyer's exit-open weights on owned hardware-remains technically and legally available. An expert that tries to become a bottleneck discovers it has only built a bridge, not a gate.

\subsection{The Price Machine}

The spot price emerges from seven forces, the way power prices emerge from load, fuel, and weather \citep{schweppe1988}:

\begin{enumerate}
  \item \textbf{Prompt difficulty}-Hard traffic is dispatched to pricier models. The mix sets how much of the ladder's expensive end the market draws on.
  \item \textbf{Output novelty}-Dense output delivers more value per metered kilobyte. Templated output prices as padding.
  \item \textbf{Task type}-Code, prose, JSON, and RAG summaries compress to systematically different densities.
  \item \textbf{Reasoning depth}-Unmetered cost the seller absorbs and folds into the rate.
  \item \textbf{Input length}-Fuel the meter never sees, likewise folded in.
  \item \textbf{Provider load}-Idle capacity is cheap, congestion expensive. Inference cannot be stored, so the price is the buffer.
  \item \textbf{Provider supply}-Every seller that clears the thresholds tightens competition. The grade floor is also the gate through which competition enters.
\end{enumerate}

Not one of these forces is visible to the buyer alone. They sit behind the credence veil. The index is what happens when an institution watches all seven continuously and publishes them as one number-the price system doing its oldest job, aggregating knowledge no single participant holds \citep{hayek1945}. That number is what the speculator will trade.

One attack deserves a mention: what economists call adverse selection \citep{rothschild1976}. One price covers easy and hard traffic alike-locked in for many hours. Hard traffic spikes, the seller serves it at a loss. When the interval ends, the price jumps to cover the harder mix. Easy traffic pays hard prices and leaves, and only hard traffic stays, priced to match. The defense is the interval-minutes against many hours. A fixed price can be beaten by anyone who knows more than the seller. A price that changes every few minutes cannot. This is not a price an expert defends-it's a tote board that reprices with the entries.

Full-requirements service makes the expert a quasi-insurer and its book a pool. Insurance runs on independence-house fires do not correlate, except at the tail-which is why a small premium can stand behind a large loss. One buyer's traffic turns hard while another's stays easy. Routed across a ladder of sellers, the deviations average out. What no pool can absorb is the systematic: a provider reprices, an architecture shifts, and every buyer is hit at once. No pool diversifies an earthquake. So the systematic is carried in the market-hedged with capacity bought forward, month for month, and borne by spread and reserves where no hedge exists. The posted rate is communal-one price per grade, fixed for its term-but the pool behind it is pari-mutuel: the index reprices with the realized mix, so the rate can go stale for no longer than one term-minutes. The expert pools the risk that averages out, hedges the risk that hits everyone at once, and prices the book in public.

\textbf{Speculators are not merely permitted in this market-they are required.} The market maker profits from the spread, not the direction. The arbitrageur closes the gap between forward curve and spot. The option trader prices the uncertainty itself-and in doing so produces what no benchmark or study ever has: a continuously market-priced measure of how fast the cost of intelligence will move, with money behind it. Compute cannot be inventoried, so speculators are also the party most financially interested in the index being real. Reputation audits the expert slowly. Competition audits it publicly. The curve audits it continuously, with capital.

Let every expert run its own exchange and publish its own index, each listing only its own graded product. Buyers need not compare grades across exchanges-that comparison belongs to arbitrageurs trading the price gap between indices. A new institution is a new attack surface: once the index settles contracts, distorting it has a payout-buying enormous inference at absurd prices near settlement, or exploiting the padding margin inherent in any volume meter. The meter alone cannot see a slightly padded answer. The expert can. It should run escalation triggers for compression anomalies concentrated near settlement windows and volume spikes from counterparties with visible derivatives positions-the signature of every index corner in history.

\section{The Assembly}

\subsection{Chapter One: The Desk}

Meter and expert in hand, the market assembles in the order every commodity market has followed: quote first, auctions second, speculation third, venue last.

The first real market is the oldest kind: instruments without a venue. Electricity ran on negotiated tariffs for decades before any pool cleared a megawatt. Chapter One's market is \textbf{the desk}-experts acting as merchants, buying upstream, quoting downstream, keeping the spread.

Wholesale is worked by hand: the desk opens accounts and negotiates rate limits-the aggregators' playbook, run with a meter and a grade. Sellers who want prices that move more often-cheap at the trough, paid at the peak-opt in to programmatic price access. Retail is the quote: one number, take it or leave it. The index, at this stage, is the desk's price sheet.

The buyer's alternative is not another market but the bill they already pay. The desk's first sale is arbitrage on that bill, and it is paid out of the waste it eliminates. Spot is the whole product: prompt in, graded kilobytes out. Certainty products exist-bilateral forwards delivered in kilobytes, full-requirements service-but with no market to hedge in, the desk writes only short, narrow contracts unless a provider signs a matching fixed-price commitment. Nothing here is a financial instrument yet.

This chapter matters for two reasons. The first is sequencing: an auction needs bidders, bidders need volume worth winning, and volume needs a retail base that trusts the quote. Auction theory is blunt: entry matters more than rules \citep{klemperer2002}. The second is hardening: the probes that will gate the auction and the triggers that will police it run in production from day one, earning a track record before money settles against them. The posted-price era is the most inefficient this market will ever be.

\subsection{Chapter Two: Auctions}

Chapter Two begins when the desk's aggregated volume grows large enough that providers compete for it in the open-the step electricity took when day-ahead pools replaced bilateral tariffs \citep{wilson2002,cramton2017}.

Two asymmetric auctions are chosen over a symmetric one, primarily for two reasons. First, a buyer purchasing raw from sellers forgoes the expert: the grading, the routing that lowers price, the warranties that catch a degrading model. Second, grading is a Red Queen expense that never amortizes, and the only way to finance it without breaking incentives is the spread-which a symmetric auction does not have.

Asymmetry has precedent on both sides. Treasury auctions are the pure one-seller case-one government selling to many banks \citep{umlauf1993}. The England and Wales electricity pool ran the mirror image for a decade: many sellers bidding against a demand forecast the market itself published \citep{green1992}. Between them they make the point: auctions succeed or fail on the pricing rule and the number of bidders, not on symmetry between the sides. What shut down the England and Wales pool was not the shape but market power: two generators controlled most of the capacity, making withholding at peak profitable and safe \citep{green1992,wolfram1999}.

The desk inherits the fixes ready-made. A uniform clearing price-every cleared trade pays the price of the last unit needed-so that bidding above the market drops you out of the running with nothing to show for it. A demand curve that slopes down, anchored to a published break-even: raise the price and the desk visibly buys less, so withholding damages the withholder \citep{hogan2014}. And an exit: open-weight models running on the buyer's own hardware.

Inference is a commodity whose clearing can be chosen rather than forced by physics: no grid imposes a batch, no warehouse imposes a curve. We borrow electricity's pooling but run it finer-every five minutes, since the batch costs nothing to repeat. There is no grid-optimization problem to solve. The auction just sorts prices. Batching makes the price expensive to manipulate: the attacker transacts at the very price they create, in public, twelve times an hour. Five minutes is the middle ground: too coarse invites corners, too fine invites noise \citep{budish2015}. The clearing price jumps when the forces jump, and stays put when someone tries to jump it.

The product the desk sells to buyers is the graded kilobyte, specified on three dimensions: grade (set by probes, never self-declared), privacy (region, zero-data-retention on or off), and delivery (a scheduled batch, or instant). An instance reads: Grade-97, EU, zero-data-retention, right now-plus quantity and price. Latency is handled by the desk: every prompt routed to the fastest qualified seller, measured in realized percentiles.

What the desk buys from sellers is raw model access on the three dimensions: model name (declared by the seller, confirmed by the probes), privacy, and delivery. Latency is measured from realized percentiles. Quantization is inferred from the probes. Sellers initially quote in input, output, and cached tokens-later, in compressed output kilobytes.

The desk aggregates its customers' traffic-realized draws plus forecast-into the demand curve it bids at the upstream auction. Probes certify each seller's live stream and do the sorting. The gap between grade clearing prices is a quality premium the market discovers, not one sellers declare. Some sellers are rival experts. Experts audit each other as a byproduct of ordinary trade. The cross-audit runs itself.

Both auctions share one engine. Asks and bids collect sealed until the batch closes. The book sorts cheapest-first, demand fills from the bottom, and every cleared trade pays the same uniform clearing price. This runs every five minutes per product. Uniform pricing does not make honest bidding rational-no rule does-but it makes shading self-punishing on either side: shade past the market and you fall out of the running entirely, with nothing to show for it \citep{wilson1979}.

Forwards clear in the same batches, in a book of their own. Enterprises bid blocks on the months their budgets live in. The desk offers schedules only as long as its hedges-capacity bought upstream, reserves behind every conversion.

Each side has its own temptation, and each its guard. Upstream, the temptation is to understate demand-bid for less than you really want, to pull the clearing price down \citep{ausubel2004}. This one dissolves on its own, because the desk is the buyer, and what it bids is its customers' real traffic, not its own money. Downstream the same trick appears in the mirror-buyers holding back orders to sink the price-but three guards wait: the desk's supply schedule slopes down, the fallback price is retail, and open-weight models on the buyer's own hardware stand as a permanent exit. And the demand curve anchored to a published break-even blunts the withholding games that plagued California \citep{borenstein2002}: raise the price, and the desk visibly buys less.

No one flips this on. The first step is to parallel-run the auction on fifteen percent of volume against the manual bilaterals. Then scale past half, publish every clearing price, switch on imbalance settlement-shortfalls buy back at the real-time price, and the penalty must exceed the best alternative revenue on a diverted kilobyte, or the market has posted a bounty on breach-and sign the first reserve contracts. When essentially all procurement runs through the auction, the index derives from clearing prices. The day outsiders take positions against the index itself, Chapter Three has begun.

\subsection{Chapter Three: Speculation}

Chapter Two ends with a number nobody prices yet: the index's future. Enterprises bid blocks for the months their budgets cover-long-only, and only against their own workload. What is missing is anyone exposed to the number itself.

Speculators hold no graded kilobytes. The product is non-storable. Even prepaid credits are claims only on service, not claims on the number. What speculators hold are claims on the number itself. The book that clears enterprise forwards already exists-the same batches, the same uniform price. Chapter Three removes the long-only clause. The desk lists forwards and caps against its published index in the same books-the market's first cash-settled instruments, settled in money rather than kilobytes-and eligible counterparties take positions, shorts included. The speculator's order differs from the enterprise's block in a single flag. Informed flow disciplines the print: a clearing price that survives people who know things is worth publishing \citep{grossman1980}. And the certainty book finds its reinsurer: the risk too big for any insurer, such as a provider repricing or an architecture shift-known as the systematic-is laid off to capital that wants exactly that exposure \citep{keynes1930,hicks1939}.

Virtual bids are financial orders that trade a price with no physical delivery behind them-borrowed from electricity, they are used to improve the market's efficiency. A virtual bid enters either book as an ordinary order and settles in cash at the printed price. Wherever a number is set administratively-the break-even anchor above all-virtual flow leans on it until it moves. Arbitrageurs measurably improved efficiency in the one market where they were studied \citep{saravia2003}, and the design carries wide precedent: PJM's capacity auction itself retired its fixed demand curve for a sloped one \citep{cramton2013}.

Trading for one's own account is unregulated everywhere. What regulation licenses is intermediation. The desk stays below the registration threshold, registers as a swap dealer, or introduces through a bank-and the counterparty gate keeps the book informed, capitalized, and few. One registration belongs to the expert itself: the first cash-settled instrument against the index in Europe makes the expert a regulated benchmark administrator, a license with duties. The paperwork precedes the first virtual bid.

Once money settles against the index, distorting it becomes a way to move money-so what keeps it honest? The classic attacks fail not because they are forbidden but because they no longer pay. A speculator settles in cash and never touches the product, so it cannot crowd delivery. A wash trade in one corner cannot move an index weighted across the market's whole volume-the node-level distortions prosecuted in electricity required isolated settlement points, and this design has none. With no clearing house yet, the speculators are the audit: their capital is staked on the index being real, and the trade that corrects a distortion is the trade that profits. Entry needs only capital and a view \citep{klemperer2002}, and their trading hardens the curve before any venue exists. In Chapter One, the desk dealt both sides of the kilobyte. Now the market has learned to hold both sides of the claim-enterprises hedging long, speculators the other side.

\subsection{Chapter Four: The Exchange}

Trading exists by now-bilaterals at the desk, schedules in the auction, positions against the index. What the venue adds is what no bilateral arrangement can manufacture: novation, anonymity, and retail. The clearing house steps between every buyer and seller. Each trade becomes two trades against it, and any position can be exited in seconds. Bilateral quotes always carried a hidden premium-compensation for being stuck with the other side-and listing dissolves it. Speculators enter because leaving is guaranteed. Liquidity is the product. The listed contract also crosses the last eligibility wall: what was wholesale-only becomes retail-reachable. The exchange doesn't license speculation. It democratizes it.

The clearing house becomes the counterparty to every trade, collects collateral, and nets offsetting positions. Collateral adjusts daily with the index's moves. The upfront deposit is sized from the index's realized volatility. Netting kills the bilateral world's costs: a seller hedging forward-sold capacity shorts the index instead of buying back from the desk, and one speculator's long January offsets another's short March without either knowing the other exists. The desk may make markets in its own index, but only matched-the ban on open exposure to its own index survives the venue.

List the months and the market grows a term structure-a curve of prices across delivery dates. No cost of storage anchors it. The curve prices expectations plus the hedging pressure of its hedgers \citep{bessembinder2002}. The calendar spread-long June, short December-is a position on when deflation arrives, since a frontier release re-marks the grades and collapses the far end faster than the near. Shorting becomes native: this market's natural drift is down, so aggregate short interest prices the rate of deflation, and the bulls are paid to argue against it. Enterprises hedge budgets years out. Sellers turn forward revenue into the collateral that finances the next training run \citep{froot1993}. And the index is audited continuously, with capital, every day rather than every dispute.

The dealer phase's caps graduate into listed options \citep{black1976}. The premium is the market's first money-priced measure of how fast the cost of intelligence will move-the variance itself, which no benchmark or study has produced. Grade re-marks become scheduled volatility events, priced like Consumer Price Index prints and earnings. A straddle-a paired bet on movement in either direction-collects whichever way the number jumps. A buyer who cannot predict prices can now buy the right not to.

Nothing rations exchanges. An exchange license is a checklist, not a quota, and rival venues listing near-identical contracts are the commodity world's ordinary state. The constraint is liquidity's network effect. The moat is the clearing house, which is where supervision should concentrate. The desk's products stay cash-settled until one shows persistent two-sided depth-standardization is earned by depth, not granted by preference-and the threat of graduation disciplines the desk's margins in the meantime.

\subsection{Where the Players Stand}

Neither side commits fully along the way. Sellers already route overflow through aggregating marketplaces-partial participation is how every commodity market began. Buyers keep the complete exit: \textbf{open weights on their own hardware.} Open weights are the market's shadow price-the option to leave constrains the expert's spread the way rooftop solar constrains the utility's tariff. The exit also clarifies who the market is for: buyers whose binding constraints are privacy or latency warranties beyond what any provider offers will self-host. Everyone else is better served paying a spread for certified quality than paying one in operational risk. For integrity, the option matters more than the exercise. And those who exit need not exit alone: open-source routers on public leaderboards suffice where no spread is charged, because the self-hoster captures less value per routing decision and cannot justify the expert's measurement complexity.

The labs may stand anywhere in this market: pure merchants selling ingredients into every expert's basket, or integrated players running their own expert arm. The lab that runs an expert arm keeps the direct customer relationship: it quotes the full-requirements rate, routes across its own ladder first, buys from rivals when they clear cheaper, and pockets the spread. Nothing in the market's integrity forbids it-the spread is paid by buyers, the grade is audited by rival experts, and a lab that grades itself generously inflates the very index its own forwards settle against. Nor does the role end at its own models: the lab that certifies open-weight traffic earns from the customers it lost, without a single price cut.

The old fear of the frontier labs-good-enough models undercutting from below-must now survive quality screening, probe audits, and reputation. The market hands them something regulation never could: a quality floor that raises the value of being the best.

\textbf{Why the seller signs.} A seller that joins gains \textbf{channels}. Enterprise customers cannot audit quality on their own, so they buy through the graded market-and every one of those deals now reaches the seller. So does the long tail of small buyers, customers too numerous to sell to one by one, who find sellers only through the index. It gains its \textbf{peak}. Today, when demand spikes, a seller throttles traffic and earns nothing extra. In a clearing market, congestion becomes a price, and hard traffic pays more for the first time. It gains its \textbf{trough}. GPUs depreciate at the same rate whether they run or sit idle, and the market is the first permanent buyer of hours that used to go unsold. It gains \textbf{reputation}. A public grade compounds: each clean audit makes the next contract easier to sell, at a smaller discount \citep{klein1981}. It gains \textbf{certainty}. Forward contracts turn volatile spot revenue into income a bank will lend against-the collateral that finances the next training run \citep{froot1993}. It gains \textbf{legitimacy}. A published grade tells the market what a sales pitch cannot. Once rivals are graded, staying out is what a withheld result looks like \citep{milgrom1981}. Sellers with spare capacity sign first. The integrated labs sign last-they already own their channels-but their baskets need rival sellers, and rival baskets hold the enterprise traffic.

\section{The State}

All this brings the market, in its final form, to the state. One might expect this to require new law. It does not. The market inherits its regulation the way electricity inherited its own: from the institutions it was always going to borrow-the meter standard and the exchange.

\subsection{First the Meter, Then the Law}

The grades began private. ``Corn No.~2 Yellow'' was an exchange-enforced standard \citep{cronon1991} for decades before the state codified it in 1916. Behind it lies older legal metrology-the Winchester bushel, the Assize of Bread-because units of trade are infrastructure, sometimes literally legislated. Electricity walked the same path twice over. The kilowatt-hour had its standards moment in the 1890s. Then, once billing depended on the meter, \textbf{the meter itself became a regulated instrument.} Utilities maintained meter-testing shops, cities appointed inspectors, and a doctored meter was fraud long before any energy statute said so. The state never defined what a kilowatt-hour meant. It certified that the ruler was straight. The sequence-\textbf{private grade, exchange supervision, statute last}-is exactly ours.

Nor is there a definitional problem. Exchanges list contracts, not commodities. Listing requires only an audited settlement number and venue supervision, which is why exchanges routinely list non-traditional underlyings. The Graded Kilobyte contract is a standard cash-settled future on a published index: the expert supplies the index, the exchange supervises the listing, traders supply the price. No institution changes its role.

This suggests a deliberate posture: \textbf{the strategy of being unremarkable.} Commodity statutes are drafted broadly-they already cover weather and bandwidth-so inference joins a lineage rather than forging a legal frontier. Fit the existing definition of a traded contract on an audited index, and inherit its supervision. Samuel Insull understood the logic a century ago, when he invited state regulation of his utilities: supervision made the tariff believable, and believable tariffs made capital cheap. An honest expert should feel the same-open books are the cheapest defense against rumor. A market requiring special understanding waits a decade for legislation. One that fits existing categories is governed the moment it lists.

\subsection{The Inherited Law}

Best of all, the exchange carries its own law. Listing inherits the oldest supervisory machinery in commercial life-anti-manipulation rules, position reporting, trade examination, record preservation-covering precisely the failure modes this market fears, on day one, at zero drafting cost. Even the sharpest fear-a trader cornering the index near expiry by buying massive inference at absurd prices-is not a novel crime but the classic squeeze, policed in grain, power, and metals for a century. The two layers reinforce each other: the expert watches compression anomalies. The exchange watches position anomalies. Neither invents its mandate.

The cautionary tale is LIBOR: a settlement index gamed by its own contributors because it rested on unaudited self-reports from interested parties. The post-LIBOR doctrine requires settlement indices to be mechanically verifiable, record-backed, and independent of those whose money depends on them. This market contains the cure in its meter: the index rests on verifiable receipts-reproducible compressed byte counts. Where grades involve judgment and cannot be receipt-verified, supervision attaches the moment the index settles contracts.

A second fear is algorithmic collusion: learning pricers can converge on supra-competitive prices without ever communicating \citep{calvano2020}. The structural defense is architectural, not supervisory: a price that is a published formula over public inputs offers nothing for an algorithm to condition on except the market itself.

\subsection{The State's Role}

Current law catches egregious fraud, not quiet bias, and small buyers remain underprotected. These gaps argue for patience, not haste. Premature laws regulate imaginary markets. Late laws regulate real ones. Over-the-counter swaps ran for decades on private rules before comprehensive statutes arrived, and electricity markets did the same-private pool rules first, federal restructuring only once the practice existed to be codified. This was not a deficiency but a research phase, letting regulators codify working private practice into universal law. If new rules are ever needed, documented failures will show exactly what to request.

An exchange layer is a regulated layer. That is how every commodity market matured, and electricity most of all. Regulation can police the venue-require the index to be mechanically derivable from published grades, firewall the routing desk from the derivatives book, open the books to examination-but it cannot police the grade, because bias is a judgment. A tampered receipt or an index input that fails to reconcile is a checkable fact, already fraud under laws decades old. The law punishes falsification. The market punishes error. And regulation closes the one door the market alone leaves open: expert--seller collusion, where a dispute between two graders returns the buyer to the exact helplessness the market was built to end-forced to pick a side on a credence question they have no tools to answer. The state cannot say which grader is honest. It can make lying checkable.

The state's whole role, then, fits within existing law: enforce contracts, punish fraud, supervise venues-with \textbf{one modest extension.} Because AI output is a credence good, the regulator's audit must reach the grading algorithm and the meter's anti-padding defenses, extending traditional oversight from books to machine mechanics. Electricity's regulators made the same extension in their day: they began by auditing ledgers and ended by testing the meters themselves. The principle was never to certify the light-only the ruler. The market that earns its exchange will find the regulation was already in the building.

\section{Conclusion}

And so the market reaches the shape the opening promised: spot prices, forward contracts, full-requirements service, and a new profession whose edge is knowing when and where to buy. Inference becomes the next commodity to learn what electricity learned in 1889: the path out of uncertainty is not a better prediction but a signature-a grade both sides can trust, a meter both sides can read, a future both sides can trade.

One caveat: all of this assumes models remain variable-different strengths, weaknesses, and prices-so that grading and routing capture real value.

The benefits compound. The market speeds the technology itself: this meter pays labs to make models cheaper, not merely better, because every caching and reasoning gain flows to margin-the next architecture is better financed. Demand compounds too. \citet{jevons1865} saw in coal that efficiency lowers price and consumption rises faster. Every reasoning gain shrinks the cost per kilobyte and enlarges the traffic to be routed, financing the expert's Red Queen rather than starving it. The measurement layer does not fear deflation-it is financed by it. And as costs fall, inference enters uses now prohibitive, from interactive media to the long tail of software that was never worth writing.

Consumers win most: certified quality at lower prices. Enterprises win with predictable budgets. Speculators win: their capital keeps the grade honest, and the law keeps their capital safe. And the news is good for the labs: every reasoning token they engineer away becomes profit funding their next architecture. Contamination-proof grading makes their lead more visible-with money behind it. Even the companies that switched to open-source models come back, paying for the labs' seal of approval on their own and others' models. And the moat they hoped regulation would give them, earned instead by the market.

A century ago the factories that built their own power plants were not wrong-they were early. The grid won because it made power boring: metered, certified, tradable, regulated-and therefore cheap. Intelligence is about to become boring in exactly the same way. That is what progress looks like from the billing department.

\section*{Disclosure}

The author is affiliated with Veltisto.ai, which is developing a commercial implementation of this proposal. Generative AI tools were used to assist with drafting and editing the prose. The author takes responsibility for the content, arguments, and any errors.

\end{document}